\documentclass[a4paper,12pt,aps,superscriptaddress]{iopart}

\usepackage{graphicx}
\usepackage{epsf}
\usepackage{iopams} 

\newcommand{\D}{\ensuremath{\mathrm{d}}}

\renewcommand{\phi}{\ensuremath{\varphi}}

\begin{document}

\title{Factorised Steady States in Mass Transport Models on an Arbitrary Graph}
\date{\today }
\author{M.\ R.\ Evans$^{1,4}$, Satya\ N.\ Majumdar$^{2}$, R.\ K.\ P.\ Zia$^{3}$ }
\address{
$^1$SUPA, School of Physics, University of Edinburgh,
Mayfield Road, Edinburgh EH9 3JZ, UK}
\address{$^2$ Laboratoire de Physique Th\'eorique et Mod\`eles Statistiques,
Universit\'e Paris-Sud, Bat 100, 91405, Orsay-Cedex, France}
\address{$^3$Department of Physics and
Center for Stochastic Processes in Science and Engineering,
Virginia Tech, Blacksburg, VA 24061-0435, USA}
\address{$^4$Isaac Newton Institute for Mathematical Sciences,
20 Clarkson Road, Cambridge, CB3 0EH, UK}

\begin{abstract}

We study a general mass transport model on an arbitrary graph consisting 
of $L$ nodes each carrying a continuous mass. The graph also has a set of 
directed links between pairs of nodes through which a stochastic portion of mass,
chosen from a site-dependent distribution,
is transported between the nodes at each time step. 
The dynamics conserves the total mass
and the system eventually reaches a steady state. 
This general model includes as special cases various previously
studied models such as the Zero-range process
and the Asymmetric random average process.   
We derive a general
condition on the stochastic mass transport rules, 
valid for arbitrary graph and for both parallel and random 
sequential dynamics, that is sufficient to guarantee that the steady state is 
factorisable. We demonstrate how this condition can be achieved in several
examples. We show that our generalized result contains as a special case
the recent results derived by Greenblatt and Lebowitz 
for $d$-dimensional hypercubic lattices with random sequential dynamics.   

\end{abstract}


\submitto{\JPA}

\maketitle

\section{Introduction}

Diverse physical phenomena such as traffic flow~\cite{traffic}, 
force propagation in granular media~\cite{Science,CLMNW}, 
clustering of buses~\cite{OEC}, aggregation and fragmentation
of clusters~\cite{MKB}, phase separation dynamics~\cite{KLMST},
shaken granular gases~\cite{MWL04,Torok} and sandpile dynamics~\cite{Jain}
share one common feature--their microscopic dynamics involves stochastic
transport of `mass', or some conserved quantity, from one point in
space to another. 
To simplify analysis, continuous space is typically replaced 
by (or ``binned'' into) discrete sites. Several such
lattice models of stochastic mass
transport have been introduced and studied, most notably the 
Zero-Range Process (ZRP)~\cite{S70,MRE00,EH05}, and the 
Asymmetric Random Average Process (ARAP)~\cite{ARAP1,ARAP2}.
These models are defined by specifying the microscopic 
dynamics, i.e. the basic stochastic rules for mass transport. Given the
dynamics, there are two principal theoretical issues: (i) to identify 
the steady state if there is any, i.e. to find the
invariant measure and (ii) once the steady state is known, to
understand various physical properties in the steady state, e.g. the
phenomenon of `condensation' that happens when a finite fraction
of the total mass condenses onto a single site~\cite{MEZ05}.   

It turns out that the step (i) itself is often very difficult
and the exact steady state is known in only very few cases~\cite{MRE00}.
In many of these known cases, the steady state 
is {\em factorisable}. This means that the steady state probability $P(\{m_i\})$ of
finding the system with mass $m_1$ at site 1, mass $m_2$ at site 2 etc is
given by a product of (scalar) factors $f(m_i)$ --- one factor for each
of the $L$ sites
of the system --- e.g. for a homogeneous system where all  sites $i$
have equivalent connectivities
\begin{equation}
P(\{m_i\})=Z(M,L)^{-1}\prod_{i=1}^L f(m_i)\;\,\delta \!\left(
\sum_{i=1}^L m_i-M\right) \;,
\label{prob}
\end{equation}
where $Z(M,L)$ is a normalisation which ensures that the integral of the
probability distribution over all configurations containing total mass $M$
is unity, hence
\begin{equation}
Z(M,L)=\prod_{i=1}^L\left[ \int_0^\infty \ensuremath{\mathrm{d}}%
m_i f(m_i)\right] \,\delta \!\left( \sum_{i=1}^L m_i-M\right) \;.  
\label{Z}
\end{equation}
Here, the $\delta $-function has been introduced to guarantee that we only
include those configurations containing mass $M$ in the integral. The
single-site weights, $f(m)$ are determined by the details of the mass
transfer rules and for a heterogeneous system may depend on the site $i$.

The advantage of having a factorisable steady state is that the step (ii)
mentioned above is often easier to carry out explicitly. This has been
demonstrated recently by an exact analysis of the condensation phenomenon
that occurs in a general class of mass transport models~\cite{MEZ05,longp}. 
This raises a natural question: when does the steady state in these mass transport 
models factorise? This issue was recently addressed in the context of
a sufficiently general `mass transport model' in one dimension, that
includes, as special cases, the previously studied ZRP, ARAP and 
the chipping model~\cite{MKB}.
In this model a mass $%
m_i$ resides at each site $i$ of a one dimensional lattice with periodic boundary conditions.
At each time, a stochastic portion, $\tilde{m}_i\le
m_i$ of the mass $m_i$ at site $i$, chosen from a distribution $\phi (\tilde{m_i}|m_i)$, 
is chipped off 
to site $i+1$. The distribution $\phi(\tilde{m}|m)$ was called the chipping kernel 
and by choosing its form appropriately one can
recover the ZRP, the ARAP and the chipping model of~\cite{MKB}.
Even though the model above is defined in discrete time where all sites
are updated in parallel, by appropriately choosing the chipping kernel
it is easy to study the continuous time limit,
which corresponds to a random sequential update sequence, as a special case~\cite{EMZ04}. 
Similarly, one can also recover, as a 
special case,
the model with only discrete masses as in ZRP. 
Thus the discrete time dynamics  generalises continuous time dynamics
but a continuous mass variable  generalises discrete mass.

A natural question, first addressed in \cite{EMZ04}, is what should be form of
the chipping kernel
$\phi (\tilde{m}|m)$ for  the the final steady state to be factorisable.
In that study, it was proved that the \emph{necessary and sufficient} 
condition for a
factorised steady state in the one dimensional
directed case  defined above is that the chipping kernel is of the form
\begin{equation}
\phi (\tilde{m}|m)= \frac{u(\tilde{m})\, v(m-\tilde{m})}{\int_0^{m}\, d{\tilde{m}}\, 
u(\tilde{m})\,v(m-\tilde{m})}
\label{phi1}
\end{equation}
where $u(z)$ and $v(z)$ are arbitrary non-negative functions. Then the
single-site weight in Eq. (\ref{prob}) is given by
\begin{equation}
f(m)=\int_0^m\, d{\tilde{m}}\, u(\tilde{m})\, v(m-\tilde{m}).   
\label{ffact}
\end{equation}
Furthermore, given a chipping kernel $\phi (\tilde{m}|m)$, sometimes it is hard
to verify explicitly that it is of the form (\ref{phi1}) and thereby to
identify the functions $u(m)$ and $v(m)$ in order to construct the weight $%
f(m)$ in Eq. (\ref{ffact}). This problem was circumvented by devising a test~\cite{ZEM04} to
check if a given explicit $\phi (\tilde{m}|m)$ satisfies the condition (\ref
{phi1}) or not. Further, if it ``passes this test,'' the weight $f(m)$ can
be found explicitly~by a simple quadrature~\cite{ZEM04}. Finally, for any
desired function $f(m)$, one can construct dynamical rules (i.e., $\phi $)
that will yield $f(m)$ in a factorised steady state. 

It was further demonstrated in Ref.~\cite{EMZ04} that the corresponding necessary and sufficient
condition in the case of random sequential dynamics in continuous time
can be easily obtained, by taking a suitable limit, from the condition for the parallel dynamics 
manifest in Eq. (\ref{phi1}). This is done by choosing the chipping kernel as 
\begin{equation} 
\phi(\tilde m|m)=
\left[1- dt\, \int_0^m \alpha(\tilde m|m)\, d{\tilde m}\right]\, \delta(\tilde m)
+ \alpha(\tilde m|m)\, dt  
\label{ct1} 
\end{equation} 
for small time increment $dt$ and $\delta(z)$ is the Dirac delta function.
The function $\alpha(\tilde m|m)$ denotes 
the `rate' at
which a mass $\tilde m$ is transferred from a site with mass $m$ to its right neighbour. Note that the 
form in Eq. (\ref{ct1}) ensures the normalization, $\int_0^m \phi(\tilde m|m)\, d{\tilde m}=1$. Then, 
the necessary and the sufficient condition for factorisable steady state, derived
from the more general condition in Eq. (\ref{phi1}), 
is that the rate $\alpha(\tilde m|m)$ must be of the following form~\cite{EMZ04}
\begin{equation} 
\alpha(\tilde m|m) = \frac{x(\tilde m)\, v(m-\tilde m)}{v(m)}, 
\label{ct2}
\end{equation} 
where $x(z)$ and $v(z)$ are arbitrary non-negative functions. The corresponding
steady state weight is then simply, $f(m)=v(m)$. 

The condition (\ref{phi1}) for factorisability in the mass transport
model was derived only in one dimension and also only for
unidirectional mass transport (from site $i$ to site $i+1$). A natural
question, therefore, is whether one can generalise this condition to
higher dimensional lattices, or to arbitrary graphs where mass
transport can take place, in general, between any pair of sites $i$
and $j$. Recently, Greenblatt and Lebowitz were able to derive a
sufficiency condition~\cite{GL} for factorisability in the mass
transport model with nearest neighbour mass transport on a regular
$d$-dimensional lattice with periodic boundary conditions, but considered
only the case of random sequential dynamics. They showed that if
$\alpha_q(\tilde m|m)\, dt$ is the probability of mass $\tilde m\le m$
being chipped off a site with mass $m$ to a nearest neighbour in the
direction $q$ (there being $2^d$ nearest neighbours on a hypercubic
lattice in $d$ dimensions), then the sufficient condition for
factorisability is a direct generalization of the condition in
Eq. (\ref{ct2}), namely that the rate function must be of the form
\begin{equation}
\alpha_q(\tilde m|m) =\frac{x_q(\tilde m) v(m-\tilde m)}{v(m)}        
\label{gl1}
\end{equation}
for each $q$,
where $x_q(z)$ for each $q$ and $v(z)$ are arbitrary functions. The steady state weight
is simple, $f(m)=v(m)$.
However, it was not possible to prove that the condition (\ref{gl1}) is also necessary~\cite{GL}, except in the case of generalized Zero-range processes.

The purpose of this paper is to derive a more general sufficiency
condition, valid for arbitrary graphs where the mass transport takes
place not necessarily between nearest neighbours and  for the more general
case of parallel
dynamics.
Our results boil down to equations (\ref{phi}) and 
(\ref{cond}). The former yields a sufficiency condition and the
latter an additional  consistency condition which must be  satisfied.

In the special case of random
sequential dynamics on a regular hypercubic lattice with nearest
neighbour mass transport, our sufficiency condition reduces to the one derived by
Greenblatt and Lebowitz. Our results, however, are considerably more
general. 

The paper is organized as follows. In Section 2, we define the mass transport
model on an arbitrary graph. In Section 3, we derive a sufficient condition
on the chipping kernels, valid for parallel dynamics on
an arbitrary graph $\cal G$, 
that would gaurantee that the steady state is factorisable. We show that
there are some additional consistency conditions that need to be satisfied
in general and we demonstrate explicitly how these conditions are satisfied
in several examples.
In Section 4, we extend our approach to random sequential dynamics on an
arbitrary graph. We conclude with a summary and discussion in Section 5.

\section{The Mass Transport Model on an Arbitrary Graph}

We consider a fixed arbitrary graph $\cal G$ consisting of $L$ nodes
labelled $i=1,2,3,\cdots, L$ and a set of directed links or channels
between certain pairs of nodes. At a given time $t$, a node $i$ has
mass $m_i\ge 0$ where $m_i$ are continuous variables. We consider
discrete time dynamics where at each step the masses at all the nodes
are updated in parallel according to the following rules. We first
define a $(L\times L)$ mass-transfer matrix $[\mu]$ as follows. An
element $\mu_{ij}$ of the matrix $[\mu]$ is identically zero at all
times if there is no directed link from site $i$ to site $j$ on $\cal
G$. If there is a directed link from $i$ to $j$, then $\mu_{ij}\ge 0$
is a non-negative stochastic variable that represents the mass
transferred from site $i$ to site $j$ during one update.
In  Fig. 1 we give 
an example which we refer to for illustrative purposes
throughout this section. The diagonal element $\mu_{ii}$ represents the mass
that stays at site $i$ at the end of the single update. We assume that
the dynamics of mass transport conserves the total mass. Thus if
$\{m_1,m_2,\dots, m_L\}$ represents the masses before the update, by
virtue of mass conservation, the row sums of the matrix $[\mu]$ are
given by, $\sum_{j} \mu_{ij}=m_i$ (see Fig. 1).  Similarly, the column
sum corresponding to a node $i$, $\sum_{j} \mu_{ji}=m_i'$ represents
the mass at site $i$ just after the update.

Note that, by definition, some elements of the matrix $[\mu]$ are
permanently zero (when there is no directed channel available for mass
transfer between a pair of sites). On the other hand, when there is an
available channel from $i$ to $j$, the matrix element $\mu_{ij}$ is a
stochastic variable which is chosen in the following way. For each
node $i$, we define a generalized `chipping kernel'
$\phi_i\left(\{\mu_{ij}\}|m_i\right)$ that represents the joint
distribution of the masses transported from site $i$ in a single
update. Here the set $\{\mu_{ij}\}$ runs over only those sites $j$
which are connected to $i$ via a directed link, and in addition it
includes the diagonal element $\mu_{ii}$. In other words,
$\{\mu_{ij}\}$ is simply the set of non-zero elements in the $i$-th
row of the mass-transfer matrix $[\mu]$.  For example, in Fig. 1 where
we have a graph of four nodes along with the directed links, we need to
define four chipping kernels as follows:
$\phi_1\left(\mu_{11},\mu_{12},\mu_{14}|m_1\right)$,
$\phi_2\left(\mu_{21}, \mu_{22}, \mu_{23}|m_2\right)$,
$\phi_3\left(\mu_{31}, \mu_{33}|m_3\right)$ and $\phi_4\left(\mu_{43},
\mu_{44}|m_4\right)$ respectively. The chipping kernels must be
normalized to unity at all sites $i$, i.e.
\begin{equation}
\int \phi_i\left(\{\mu_{ij}\}|m_i\right)\, \delta\left(\sum_{j}\mu_{ij}-m_i\right)\, \prod_{j} 
d\mu_{ij}=1
\label{normphi1}
\end{equation}
where it is implied that 
the index $j$ in the sum as well as in the product runs over all
the sites that $i$ feeds into, including $i$ itself.  Thus in this
model  the amount of mass transported from a
given site $i$ in one update depends only on that site $i$, and not,
e.g. on the destination sites to which the mass is transported. Also, we
assume that the chipping kernels $\phi_i\left(\{\mu_{ij}\}|m_i\right)$
do not contain the time $t$ explicitly.  Note also that this chipping
kernel $\phi_i\left(\{\mu_{ij}\}|m_i\right)$ generalises
that of \cite{EMZ04,ZEM04} sufficiently  to
include hypercubic lattices\cite{GL}  
or more complicated graphs.

\begin{figure} 
\includegraphics[width=4in]{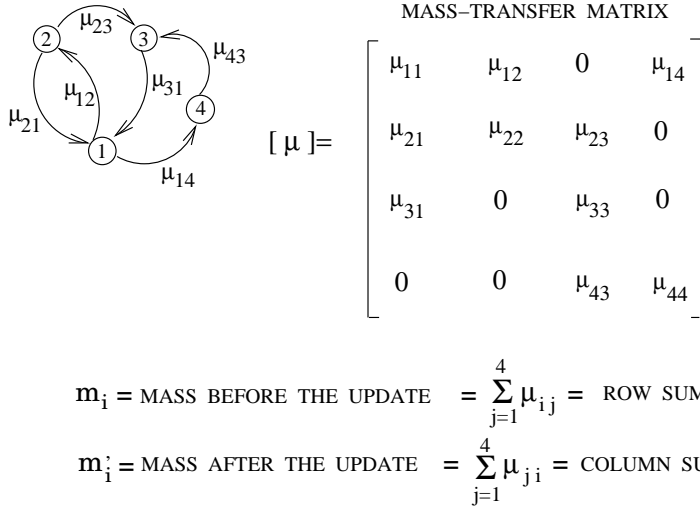} 
\caption{\label{fig:graph1} An example  graph with four nodes labelled $1$, $2$, $3$ and $4$
and directed links between certain pairs of nodes. The associated
$(4\times 4)$ mass-transfer matrix $[\mu]$ is shown, whose element $\mu_{ij}$ denotes
the stochastic mass transferred from site $i$ to site $j$ in one single update, provided
there is a directed link between the two sites. If there is no directed link, the corresponding
matrix element is always identically zero. 
The diagonal element $\mu_{ii}$ is the amount of mass that stays at site $i$ during
the update. 
The row sum and the column sum associated with any node 
$i$,  $\sum_{j} \mu_{ij}=m_i$ and $\sum_{j}\mu_{ji}=m_i'$, represent respectively
the mass at $i$ before and after the update.}
\end{figure}

The chipping kernels thus specify the dynamics, i.e. the mass update
rules. Given these kernels, we next ask what is the steady state joint
probability distribution of masses $\underline{m}\equiv
\{m_1,m_2,\cdots, m_L\}$, i.e.  $P\left(\underline{m}, t\to
\infty\right)$. In particular, our goal is to determine
the  properties
of  the chipping kernels $\phi_i\left(\{\mu_{ij}\}|m_i\right)$
required in order to guarantee that the steady state joint probability
distribution is factorisable, i.e. of the form
\begin{equation}
P\left(\underline{m}, t\to \infty\right) = {Z(M,L)}^{-1}\, \left[\prod_{i=1}^L f_i(m_i)\right]\,
\delta\left(\sum_{i=1}^L m_i -M\right)
\label{gpm1}
\end{equation}
where the normalization constant is given by
\begin{equation}
Z(M,L)= \prod_{i=i}^L \left[\int_0^{\infty} dm_i f_i(m_i)\right]\, \delta \left(\sum_{i=1}^L m_i 
-M\right).
\label{gpt1}
\end{equation}
Besides, if the steady state factorises as in Eq. (\ref{gpm1}), we
would also like to know the single-site weights $f_i(m_i)$ in terms of
the prescribed chipping kernels
$\phi_i\left(\{\mu_{ij}\}|m_i\right)$. Note that on an arbitrary graph
$\cal G$, the single-site weights $f_i(m_i)$ are, in general,
different from site to site. Hence there is an additional subscript
$i$ in $f_i(m_i)$. On a homogeneous graph, where all sites have
equivalent connectivities, the weight function $f(m)$ does not depend
on the site $i$ explicitly as in Eq. (\ref{prob}).

\section{Factorisable Steady State on an Arbitrary Graph}  

Since the dynamics conserves the total mass, at any time $t$ we can write
$P\left(\underline{m}, t\right)\propto F(\underline{m}, t) \delta\left(\sum_i {m_i}-M\right)$
where $M$ is the total mass and $F(\underline{m}, t)$ is the unnormalized weight
at time $t$. Below, we will first write down the general evolution equation of
the weight $F(\underline{m}, t)$ under the mass transport rules prescribed 
by the chipping kernels. While the notations that we will use for a general
graph $\cal G$ may seem a bit complicated, it is instructive to keep the simple example 
in Fig. 1 in mind and use it as a guide to the general notations.

Let us consider a single update from time $t$ to time $t+1$. Let $\underline{m}\equiv 
\{m_1, m_2, \cdots, m_L\}$ denote the masses at time $t$ before the update and
$\underline{m'}\equiv \{m_1',m_2',\cdots, m_L'\}$ denote the masses at time $t+1$
after the update. In terms of the elements of the mass-transfer matrix $[\mu]$, 
we thus have, $m_i = \sum_{k} \mu_{ik}$ and
$m_i'= \sum_{k} \mu_{ki}$ for all $i$. The master equation for the evolution of the weight
then reads
\begin{eqnarray}
\lefteqn{F(\underline{m'}, t+1)=}\label{Meq}\\
&&  \prod_{i=1}^L
\left[ \int_0^\infty \D m_i \prod_j \int \D \mu_{ij}
\phi_i \left( \{ \mu_{ij}\} | m_i\right)
 \delta(m_i'- \sum_k \mu_{ki})
 \delta(m_i- \sum_k \mu_{ik}) \right] F(\underline{m},t)
\nonumber
\end{eqnarray}
where the product over $j$ runs over only the sites to which the site $i$ feeds into, i.e.,
when there is a directed link between sites $i$ and $j$ (note that this set includes
the site $i$ itself). For other sites 
that are not
connected to $i$ by a directed link, the corresponding matrix element $\mu_{ij}=0$ identically (see
Fig. 1) and hence they are not integration variables in Eq. (\ref{Meq}).

Our strategy now is to assume that the steady state factorises and to
determine a sufficent condition for this assumption to hold.
First, we take the $t\to \infty$ limit on both sides of Eq. (\ref{Meq}) and assume
that the steady state weight factorises
\begin{equation}
F(\underline{m})=\prod_i f_i(m_i)
\label{Ffac}
\end{equation}
and write
\begin{equation}
f_i(m_i) \phi_i \left( \{ \mu_{ij}\} | m_i\right)
 = {\cal P}_i (\{\mu_{ij}\})\delta(m_i-\sum_{j}\mu_{ij})\;,
\label{phimun}
\end{equation}
thus
\begin{equation}
\prod_{i=1}^L f_i(m_i')=  \prod_{i=1}^L
\left[ \int_0^\infty \D m_i \prod_j \int  \D \mu_{ij}
{\cal P}_i (\{\mu_{ij}\})
 \delta(m_i'- \sum_k \mu_{ki})
 \delta(m_i- \sum_k \mu_{ik}) \right] \;.
\label{Meq2}
\end{equation}
Now we Laplace transform this equation to obtain
\begin{equation}
\prod_{i=1}^L g_i(s_i)= \prod_{i=1}^L \left[ \int_0^\infty \D m_i \prod_j \int \D \mu_{ij}
{\cal P}_i (\{\mu_{ij}\})\delta(m_i- \sum_k \mu_{ik})\,{\rm e}^{-s_i \sum_{k} \mu_{ki}}
\right].
\label{Meq2a}
\end{equation}
where we have defined $g_i(s_i)= \int_0^{\infty} f_i(m) e^{-s_i\, m} dm$. 
Next we trivially perform the integration over the $m_i$ variables on the rhs
of Eq. (\ref{Meq2a}) and rearrange the $s_i$ to get
\begin{equation}
\prod_{i=1}^L g_i(s_i)=
\prod_{i=1}^L \left[   \prod_j \int \D \mu_{ij}
{\cal P}_i (\{\mu_{ij}\})
\rme^{-\sum_k s_k \mu_{ik}} \right]
\equiv
\prod_i X_i(\{s_k\}_i )\;.
\label{LT}
\end{equation}
where  
$\{s_k\}_i$ indicates that $k$ runs over the sites to which $i$ is connected
by a directed link, including the site $i$ itself. For example, for the graph 
in Fig. 1, 
we have
\begin{equation}
g_1(s_1) g_2(s_2) g_3(s_3) g_4(s_4) = X_1(s_1,s_2,s_4)\, X_2(s_1,s_2,s_3)\, X_3(s_1,s_3)\, X_4(s_3,s_4).
\label{LTfig1}
\end{equation}

One solution (not the most general as we shall discuss in 
section \ref{sec:conc} ) to (\ref{LT}) is
\begin{equation}
X_i( \{s_k\}_i)
 =\prod_j K_{ij}(s_j)
\label{ansatz}
\end{equation}
so that 
\begin{equation}
g_i(s_i)
= \prod_j K_{ji}(s_i)\;.
\label{g}
\end{equation}
As before, the product in Eq. (\ref{ansatz}) runs over only those sites $j$ that the site $i$ feeds 
into (including itself),
whereas in Eq. (\ref{g}) the product runs over only those sites $j$ that feed onto $i$ (including
the site $i$).
For example, for 
the graph in Fig. 1, we will have 
\begin{eqnarray}
X_1(s_1,s_2,s_4) &= & K_{11}(s_1)\, K_{12}(s_2)\, K_{14}(s_4) \nonumber \\
X_2(s_1,s_2,s_3) &= & K_{21}(s_1)\, K_{22}(s_2)\, K_{23}(s_3) \nonumber \\
X_3(s_1,s_3) &= & K_{31}(s_1)\, K_{33}(s_3) \nonumber \\
X_4(s_3, s_4) &=& K_{43}(s_3)\, K_{44}(s_4)
\label{ans1fig1}
\end{eqnarray}
and correspondingly 
\begin{eqnarray}
g_1(s_1)&=& K_{11}(s_1)\, K_{21}(s_1)\, K_{31}(s_1) \nonumber \\
g_2(s_2) &=&  K_{12}(s_2)\, K_{22}(s_2) \nonumber \\
g_3(s_3) &=& K_{23}(s_3)\, K_{33}(s_3)\, K_{43}(s_3) \nonumber \\
g_4(s_4) &=& K_{14}(s_4) K_{44}(s_4).
\label{ans2fig1}
\end{eqnarray}

Since the rhs of Eq. (\ref{g}) is a product of Laplace transforms, its 
inverse Laplace transform $f_i(m_i)$ therefore has the convolution form
\begin{equation}
f_i(m)= 
\left[\prod_{*j} v_{ji}\right] (m)\;
\label{f}
\end{equation}
where the notation indicates a product
over sites $j$ feeding into $i$ which is a multiple convolution and 
the function $v_{ji}(m)$ is the inverse Laplace transform of 
$K_{ji}(s_i)$, i.e.
\begin{equation}
K_{ji}(s_i)=\int_0^{\infty} v_{ji}(\sigma)\, {\rm e}^{-s_i \sigma} d\sigma 
\label{invlt1}
\end{equation}
Again, going back to Fig. 1, we will have, for example, by inverting Eq. (\ref{ans2fig1})
the following convolution for the site labelled $1$
\begin{equation}
f_1(m_1)= \int v_{11}(m_1-\sigma_{21}-\sigma_{31})\, v_{21}(\sigma_{21})\, v_{31}(\sigma_{31}) 
d\sigma_{21}\, 
d\sigma_{31}  
\label{invlt2}
\end{equation}
and similarly for the sites labelled $2$, $3$ and $4$.

Having obtained the form of $f_i(m)$, let us ask what does this imply for the
chipping kernels. To see that, we go back to the definitions of $X_i$ in Eq. (\ref{LT}).
Substituting the ansatz for $X_i$ in Eq. (\ref{ansatz}) on the rhs of Eq. (\ref{LT})
we see that for each $i$
\begin{equation}
\prod_j \int \D \mu_{ij}
{\cal P}_i (\{\mu_{ij}\})
{\rm e}^{-\sum_k s_k \mu_{ik}}
=\prod_j K_{ij}(s_j)
\label{LT2}
\end{equation}
For example, for the site $1$ in Fig. 1, we have
\begin{equation}
\int  d\mu_{11}\, d\mu_{12}\, d\mu_{14}\,
P_1(\mu_{11},\mu_{12},\mu_{14})\, {\rm e}^{-s_1\, \mu_{11}-s_2\,\mu_{12} 
-s_4\, \mu_{14}} 
= K_{11}(s_1)\, K_{12}(s_2)\, K_{14}(s_4),
\label{LT2fig1}
\end{equation}
and similarly for the sites $2$, $3$ and $4$. Eq. (\ref{LT2}) immediately implies for the general case
\begin{equation}
{\cal P}_i (\{\mu_{ij}\})
= \prod_j v_{ij}(\mu_{ij})
\label{prodP1}
\end{equation}
where $j$ runs over the sites which site $i$ feeds into
and  $v_{ij}$ is the Laplace inversion of $K_{ij}$ as in Eq. (\ref{invlt1}). For example, for the site
$1$ in Fig. 1 we have
\begin{equation}
P_1(\mu_{11},\mu_{12},\mu_{14}) = v_{11}(\mu_{11})\, v_{12}(\mu_{12})\, v_{14}(\mu_{14}).
\label{prodPfig1}
\end{equation}
Substituting Eq. (\ref{prodP1}) in Eq. (\ref{phimun}) we get the following expression
for the chipping kernel
\begin{equation}
\phi_i \left( \{ \mu_{ij}\} | m_i\right)
= \frac{ \prod_j v_{ij}(\mu_{ij}) }{ \left[ \prod_{*k} v_{ki}\right](m_i)}\;.
\label{phi}
\end{equation}
In the numerator the product  is over sites $j$ which site $i$
feeds into via a directed link connecting $i$ to $j$ (this includes
site $i$ itself), and
one has the constraint $\sum_{j} \mu_{ij}= m_i$. 
In the denominator, on the other hand,  the product is over sites $k$ feeding into site $i$
(including once again the site $i$ itself).
Going back again to the example in Fig. 1, we have for the site $1$ the following chipping kernel
\begin{equation}
\phi_1(\mu_{11}, \mu_{12}, \mu_{14}|m_1)= 
\frac{ v_{11}(\mu_{11})\, v_{12}(\mu_{12})\, v_{14}(\mu_{14})}{\int v_{11}(m_1-\sigma_{21}-\sigma_{31})\, 
v_{21}(\sigma_{21})\, 
v_{31}(\sigma_{31}) d\sigma_{21}\, d\sigma_{31}}
\label{phifig1}
\end{equation}
where $\mu_{11}+ \mu_{12} + \mu_{14} = m_1$. Similarly one can easily express the chipping kernels for the 
other sites $2$, $3$ and $4$ in terms of the functions $v_{ij}$.

However this is not quite the end of the story since the chipping kernel
$\phi_i(\{\mu_{ij}\}|m_i)$ must obey a key consistency condition, namely 
its normalization: Eq. (\ref{normphi1}).
This implies from
Eq. (\ref{phi}) that for each site $i$ 
\begin{equation}
\left[ \prod_{*j}v_{ji}\right](m) =
\left[ \prod_{*j}v_{ij} \right](m)
\label{cond}
\end{equation}
Again, the convolution on the lhs is over sites $j$ feeding into site $i$
but the convolution on the rhs is over sites $j$ which $i$ feeds into.
For example, for the site $1$ in Fig. 1, this consistency condition reads
\begin{eqnarray}
\lefteqn{
\int v_{11}(m_1-\sigma_{21}-\sigma_{31})\, v_{21}(\sigma_{21})\,
v_{31}(\sigma_{31})\,d\sigma_{21}\,
d\sigma_{31} =} \nonumber \\ 
&& \int v_{11}(m_1-\mu_{12}-\mu_{14})\, v_{12}(\mu_{12})\, v_{14}(\mu_{14})\,
d\mu_{12}\, d\mu_{14}.
\label{consfig1}
\end{eqnarray}
One can write down similar consistency conditions for sites $2$, $3$ and $4$ also.

In summary, a sufficient  condition for factorisation is that the chipping kernels 
are product form as in Eq. (\ref{phi}) where $v_{ij}$ are
non-negative functions that satisfy the consistency conditions in
Eq. (\ref{cond}), but otherwise are arbitrary.  If this condition holds,
then the steady state will factorise with single-site weight $f_i(m)$
given by the convolution formula in Eq. (\ref{f}).  These conditions form
the central result of this paper.
The conditions arise from making the ansatz in Eq. (\ref{ansatz})
which immediately  implies
Eq. (\ref{f}) for $f_i(m_i)$
and Eq.(\ref{prodP1}) for 
${\cal P}_i (\{\mu_{ij}\})$. These two equations, when substituted in Eq. (\ref{phimun}),
gives Eq. (\ref{phi}) for the chipping kernel. Finally, 
Eq. (\ref{cond}) arises from the requirement that the chipping kernel
in Eq. (\ref{phi}) is normalized to unity.

Condition (\ref{phi},\ref{cond}) is certainly a sufficient
condition. Whether it is also a necessary condition remains a
non-trivial issue. In the special case of one dimensional graph with
unidirectional transport, it was explicitly proved in
Ref.~\cite{EMZ04} that this is also a necessary condition. Similarly,
this is true (see below) in the case of a complete graph where every
site is connected to every other site.  However, it is unlikely, in general,
that (\ref{phi}) is also necessary. In section~\ref{sec:conc}, 
we provide an
explicit example showing solutions to (\ref{LT}) which are not of the form
(\ref{phi}). Although we are unable, for a technical reason to be 
discussed later,
to prove that this class of solutions are also ``legitimate,'' we feel
that valid solutions different from (\ref{phi}) should exist. Beyond this
simple example, finding the most general valid solution to Eq. (\ref{LT})
for the arbitrary graph seems to pose significant challenges.

\subsection{Specific Examples}
\label{sec:spec}
Let us now discuss several special cases where one can directly verify that 
the consistency conditions in Eq. (\ref{cond}) are satisfied automatically by
the functions $v_{ij}$ due to the nature of the underlying graph $\cal G$, 
thereby guaranteeing factorisability provided the chipping kernels are
of the product form as in Eq. (\ref{phi}).     

\vspace{0.4cm}

\noindent{\bf Example 1: Mass transport in one dimension over extended
range:} Consider a one dimensional latice with periodic boundary
conditions where mass from a given site can get transported up to a
range $l$ to the left or to the right. 
In this case
\begin{equation}
\phi_i \left( \{ \mu_{ij}\} | m_i\right)
\phi_i \left( \mu_{i,j-l}\ldots \mu_{i,j+l} | m_i\right)
\end{equation} and we have a translationally invariant chipping kernel
i.e. $\phi_i(\cdot|m)$ is the same function of $(2l+1)$ arguments for each site.
Then the consistency condition
(\ref{cond}) reads
\begin{equation}
[v_{i,i-l}*v_{i,i-l+1}*\cdots* v_{i,i+l-1}*v_{i,i+l}](m_i)=[v_{i-l,i}*v_{i-l+1,i}*\cdots*v_{i+l-1,i}
*v_{i+l,i}](m_i)
\label{1dxr}
\end{equation}
which is automatically satisfied due to the translational invariance $v_{i,i+d}(\mu)=v_{i-d,i}(\mu)$.

\vspace{0.4cm}

\noindent{\bf Example 2: Symmetric chipping rule:} It is easy to see that if the chipping rules are
symmetric, i.e. $v_{ij}(\mu)=v_{ji}(\mu)$, then the consistency condition (\ref{cond})
holds automatically.

\vspace{0.4cm}

\noindent{\bf Example 3: A more general chipping rule:} A more general way to
meet the condition (\ref{cond}) is to assume that the chipping rules
are such that at each site $i$, for every directed link $(i\to j)$ out
of site $i$, there is an incoming directed link $(k\to i)$ to $i$ from
some other site $k$ with $v_{ij}(\mu)= v_{ki}(\mu)$. For example, in
the $1$-d example above, this is achieved through $v_{i,i+d}(\mu)=
v_{i-d,i}(\mu)$. Similarly, one can achieve this on a hypercubic
lattice in arbitrary dimensions and on any graph where the number
of links out of a site equals the number of links into it.

\vspace{0.4cm}

\noindent{\bf Example 4: Complete graph
with permutationally invariant chipping kernel:} So far we have proved that
Eq. (\ref{phi}) is a sufficient condition for factorisability on any
arbitrary graph provided we can satisfy the consistency condition
(\ref{cond}) for each site. As discussed before, to prove that this
condition is also necessary seems hard for a general graph. However,
one can prove this for a complete graph where every site is connected
to every other site via a directed link
{\em and} the chipping kernel is permutationally invariant. This can be proved as
follows.  On a complete homogeneous graph the functions $g_i(s)=g(s)$
in Eq. (\ref{LT}) do not depend explicitly on the site index
$i$. Moreover, since every site is connected to every other sites
and using the fact that the chipping kernel is permutationally invariant
Eq. (\ref{LT}) simply becomes
\begin{equation}
g(s_1)g(s_2)\cdots g(s_L)= {\left[ X(s_1,s_2,\cdots, s_L)\right]}^L.
\label{cg1}
\end{equation}
Clearly $X(s_1,s_2,\cdots,s_L)= K(s_1)K(s_2)\cdots K(s_L)$ with $K(s)$ arbitrary, is {\em one} 
solution
to Eq. (\ref{cg1}) and thereby $g(s)=[K(s)]^L$. As in the general case, this will then lead 
to the sufficiency condition (\ref{phi}). To prove that $X(s_1,s_2,\cdots,s_L)= 
K(s_1)K(s_2)\cdots 
K(s_L)$ is also the most general form of the solution that one can write down for Eq. 
(\ref{cg1}),
we take logarithm on both sides of Eq. (\ref{cg1}) and then take derivatives with respect to $s_i$
and $s_j$ with $i\ne j$. This gives
\begin{equation}        
\frac{\partial^2 \ln X}{\partial s_i \partial s_j}=0,
\label{cg2}
\end{equation}
for any $i\ne j$. It is then easy to see that the most general solution of the partial differential 
equation (\ref{cg2}) is of the form, 
$X(s_1,s_2,\cdots,s_L)= K_1(s_1)\,K_2(s_2)\cdots K_L(s_L)$
where $K_i(s)$ are arbitrary functions. Since the graph is homogeneous, we also have
$K_i(s)=K(s)$ independent on the site index $i$. Thus 
$X(s_1,s_2,\cdots,s_L)= K(s_1)\,K(s_2)\cdots K(s_L)$ with $K(s)$ being an arbitrary function,
is the only solution of Eq. (\ref{cg1}) that respects homogeneity. Since this solution
finally leads to the sufficiency condition, the uniqueness of its form
guarantees then that Eq. (\ref{phi}) is both necessary and sufficient. 
Note that the consistency condition (\ref{cond})
is automatically satisfied in this case.

\section{Random Sequential Dynamics}

The sufficiency condition in Eq. (\ref{phi}) and the associated
consistency condition in Eq. (\ref{cond}), derived above for parallel
update dynamics in discrete time, can be easily extended to the case
of random sequential dynamics. This can be achieved by letting the
probability of the chipping event in a time step $\propto dt$ so that,
to leading order in $dt$ for small $dt$, at most one chipping event can
occur in the whole graph $\cal G$ per update, i.e.  the chipping
events occur sequentially one per update. In addition, taking $dt\to
0$ one can obtain the continuous time limit where chipping events
occur with `rates' per unit time. Thus, the corresponding sufficiency
condition for the random sequential dynamics will be specified in
terms of the chipping `rate' kernels, rather than the chipping
probablity kernels $\phi_i$ in parallel dynamics. To translate the
sufficiency condition in Eq. (\ref{phi}) valid for the probability
kernels into one that is valid for `rate' kernels, we first redefine
the functions $v_{ij}(\mu_{ij})$, for all $i\ne j$, in the following
way
\begin{equation}
v_{ij}(\mu_{ij})= \delta(\mu_{ij}) + x_{ij}(\mu_{ij}) dt
\label{vrs1}
\end{equation}   
where $x_{ij}(\mu_{ij})$ are arbitrary functions. The diagonal functions $v_{ii}(\mu_{ii})$
are left unchanged. 

With this re-definition of $v_{ij}$, the steady state weight in
Eq. (\ref{f}) becomes
\begin{equation}
f_i(m) = v_{ii}(m) + dt\, \left[\sum_{j\ne i} \int_0^m v_{ii}(m-\mu_{ji})\, x_{ji}(\mu_{ji})\, d\mu_{ji} 
\right] + O(dt^2)   
\label{wrs1}
\end{equation}
where the sum over $j$ in the second term runs over all sites $j\ne i$ that feed
into site $i$ on $\cal G$. Using the re-defined  
$v_{ij}$ in Eq. (\ref{vrs1}) one can similarly rewrite the chipping kernels
in Eq. (\ref{phi}). Since the diagonal elements play a special role, it is convenient
to redefine the chipping kernel only in terms of non-diagonal elements, i.e.,
$\phi_{i}\left(\{\mu_{ij}\}|m_i\right)\equiv \phi_i\left(\{\mu_{ij}\}'|m_i\right)$ where
$\{\mu_{ij}\}'$ denotes the set of matrix elements in the row $i$ without
the diagonal element $\mu_{ii}$. We are allowed to get rid of the diagonal element
using the row sum, $\sum_{j} \mu_{ij} = m_i$. For example, for the 
graph in Fig. 1, we will
rewrite, $\phi_1(\mu_{11}, \mu_{12}, \mu_{14}|m_1)\equiv \phi_1(\mu_{12}, \mu_{14}|m_1)$.
Substituting Eq. (\ref{vrs1}) in Eq. (\ref{phi}) and taking 
the limit $dt\to 0$, one gets
\begin{eqnarray}
\phi_i\left(\{\mu_{ij}\}'|m_i\right) &=& 
\left[1- \frac{dt}{v_{ii}(m_i)}\, \sum_{j\ne i}
[x_{ji}*v_{ii}](m_i)\right]\,  \prod_{j\ne i} \delta(\mu_{ij}) + \nonumber \\
&+& dt\, \left[\sum_{j\ne i} 
\frac{x_{ij}(\mu_{ij})\, v_{ii}(m_i-\mu_{ij})}{v_{ii}(m_i)}
\prod_{k\ne j}\delta(\mu_{ik})\right] + O({dt}^2) 
\label{phirs1}
\end{eqnarray}
where $[x*y](m)= \int_0^m x(\sigma) y(m-\sigma) d\sigma$ denotes the convolution integral.  
The notation in Eq. (\ref{phirs1}) may look a bit complicated, but actually 
it's rather simple.
For example, for the graph in Fig. 1, the kernel in Eq. (\ref{phirs1}) for the site labelled 
$1$ reads,

\begin{eqnarray}
\lefteqn{\phi_1(\mu_{12}, \mu_{14}|m_1)=}\label{phirsfig1}
\\
&&
 \left[1- \frac{dt}{v_{11}(m_1)}\left( [v_{11}*x_{31}](m_1)
+[v_{11}*x_{21}](m_1)\right)\right]\, \delta(\mu_{12})\,\delta(\mu_{14}) + \nonumber\\
&+& dt\, \left[\frac{x_{12}(\mu_{12}) v_{11}(m_1-\mu_{12})}{v_{11}(m_1)}\,\delta(\mu_{14})
+ \frac{x_{14}(\mu_{14}) v_{11}(m_1-\mu_{14})}{v_{11}(m_1)}\,\delta(\mu_{12})\right]. 
\nonumber
\end{eqnarray}

Taking $dt\to 0$ limit in Eq. (\ref{phirs1}) we then obtain the
continuous time limit where a portion $\mu_{ij}$ gets chipped off the
site $i$ with mass $m_i$, to another site $j$ which is connected to
$i$ by a directed link from $i$ to $j$ on $\cal G$, with a rate
$x_{ij}(\mu_{ij})v_{ii}(m_i-\mu_{ij})/{v_{ii}(m_i)}$. Also, we notice
that in the limit $dt\to 0$, the steady state weight $f_i(m_i)$ is
given simply from Eq. (\ref{wrs1})
\begin{equation}
f_i(m_i)= v_{ii}(m_i).
\label{wrs2}
\end{equation}

However, as in the case of parallel update, we need the functions $x_{ij}$ to satisfy
certain additional consistency conditions. This is because the chipping kernel
$\phi_i\left(\{\mu_{ij}\}'|m_i\right)$, when integrated over its arguments $\{\mu_{ij}\}'$
must give unity. Integrating Eq. (\ref{phirs1}) then gives the required consistency condition
that must be satisfied for each node $i$, 
\begin{equation}
\sum_{j\ne i} [x_{ij}*v_{ii}](m_i) = \sum_{j\ne i} [x_{ji}*v_{ii}](m_i),
\label{consrs1}
\end{equation}
where the sum on the lhs runs over all sites $j$ that the site $i$ feeds into
(excluding the site $i$ itself), but on the rhs the sum runs over all sites $j$
that feed into site $i$ (excluding the site $i$ itself). As an example of this condition
for the site labelled $1$ in Fig. 1, we have
\begin{equation}
[x_{12}*v_{11}](m_1) + [x_{14}*v_{11}](m_1)= [x_{21}*v_{11}](m_1)+ [x_{31}*v_{11}](m_1).
\label{consrsfig1}
\end{equation}
Similar consistency conditions can be wriiten down for sites $2$, $3$ and $4$ in Fig. 1.

Thus, in summary, for the mass transport model on an arbitrary graph $\cal G$ with 
random sequential dynamics specified by the rates $\alpha_{ij}(\mu|m_i)$ 
of mass $\mu$ to be transferred from site $i$ to site $j$ (provided there
is a directed link between $i$ and $j$ on $\cal G$), a sufficiency condition
for the steady state to be factorisable with weights $f_i(m_i)$ is that
$\alpha_{ij}(\mu|m_i)$ is of the form
\begin{equation}
\alpha_{ij}(\mu|m_i)= \frac{x_{ij}(\mu) v_{ii}(m_i-\mu)}{v_{ii}(m_i)}   
\label{condf1}
\end{equation}
for all $i$, where $x_{ij}$ and $v_{ii}$ are functions that must satisfy
the consistency conditions in Eq. (\ref{consrs1}), but otherwise arbitrary.
Also, the corresponding steady state weight is then given simply by 
$f_i(m_i)=v_{ii}(m_i)$.
Note that though formally Eq. (\ref{condf1}) looks like a direct generalization 
of the one dimensional condition in Eq. (\ref{ct2}), the additional consistency
conditions in Eq. (\ref{consrs1}) are nontrivial to satisfy. For the one dimensional
example in Eq. (\ref{ct2}), the corresponding consistency condition, 
$[x_{i,i+1}*v_i](m_i)=[x_{i-1,i}*v_i](m_i)$
is automatically satisfied due to the translational invariance, $x_{i-1,i}(\mu)=x_{i,i+1}(\mu)$.

\subsection{Specific example of hypercubic lattice}
It is easy to verify that the consistency condition in Eq. (\ref{consrs1})
is automatically satisfied 
in the four example cases of section~\ref{sec:spec} 
(as it should be since  random sequential dynamics is just a 
limit of the discrete time case).
Let us do this explicitly for the case
where the graph $\cal G$ is a homogeneous hypercubic lattice with
periodic boundary conditions  and
mass transfer takes place only between nearest neighbours. 
For a hypercubic lattice,
from each site $i$ there are $2^d$ outgoing links to the $2^d$ nearest neighbours of
$i$. Similarly, there are 
exactly $2^d$ incoming links to site $i$ from its nearest neighbours. 
Then the sufficiency condition in Eq. (\ref{condf1}) can be written in
a simplified notation 
\begin{equation}
\alpha_{i,q}(\mu|m_i)= \frac{x_{i,q}(\mu)v_{ii}(m_i-\mu)}{v_{ii}(m_i)}
\label{condcubic1}
\end{equation} 
where the index $q=\pm 1$ runs over the $2^d$ directions $\{\pm e_1, \pm e_2, \cdots, \pm e_d\}$,
$\alpha_{i,q}(\mu|m_i)$ denotes the rate of transfer of mass $\mu$ from site $i$ with mass $m_i$ 
in the direction $q$ and
$x_{i,q}$ denotes the function associated with the link $(i, i+q)$. Similarly, the
the consistency condition
in Eq. (\ref{consrs1}) can be rewritten as
\begin{equation}
\sum_q [x_{i,q}*v_{ii}](m_i) = \sum_{j\in {\rm neighbours\,\,\, of}\,\,\, i} [x_{j,-q}*v_{ii}](m_i).
\label{conscubic}
\end{equation}

We next use the fact that the lattice is homogeneous, i.e. it is translationally
invariant in all directions. Clearly then $x_{i-q,q}(z)=x_{i,i+q}=g_q(z)$ due to translational 
invariance in the $q$-th direction. In that case the condition in Eq. (\ref{conscubic}) is clearly
satisfied at all $i$. Also, due to the translational invariance, the rate function
$\alpha_{i,q}(\mu|m_i)=\alpha_q(\mu|m_i)$ does not depend on the site index $i$.
By the same requirement,
$v_{ii}(m_i)=v(m_i)$.
Thus, the sufficiency
condition in Eq. (\ref{condf1}) simply reads,
\begin{equation}
\alpha_{q}(\mu|m_i)= \frac{g_q(\mu) v(m_i-\mu)}{v(m_i)}
\label{conscubic2}
\end{equation}
where $g_q(z)$ and $v(z)$ are arbitrary non-negative functions. The consistency
conditions are automatically satisfied as proved above. The steady state single
site weight $f(m_i)$ is simply, $f(m_i)=v(m_i)$, and naturally it does not
depend on the site index $i$ explicitly. The condition in Eq. (\ref{conscubic2})
is precisely that derived by Greenblatt and Lebowitz. Thus, our general 
sufficiency condition, valid for an aribtrary graph $\cal G$, recovers 
this special case when $\cal G$ is a homogeneous hypercubic lattice
with nearest neighbour mass transport.

\subsection{ZRP on arbitrary graph}
We can also check that the steady state for the continuous time
zero-range process on
an arbitrary graph is recovered.  The zero-range process involves 
discrete masses and is specified by the rates  for a
a unit of mass to hop from
site $i$ to $j$. In the case where these rates are given by
\begin{equation} \alpha_{ij}(1|m_i) = y_{ij} w_i(m_i)\; ,
\label{zrpag}
\end{equation}
where $w_i(m_i)$ is the total rate for the unit mass leaving site
$i$ and $y_{ij}$ is the probability that the random destination for a hop from site $i$ is $j$,
the steady state factorises with single-site weight
\begin{equation}
f_i(m_i) = \frac{ p_i^{m_i}}{ \prod_{n=1}^{m_i} w_i(n)}
\label{zrpagf}
\end{equation}
where $p_i = \sum_{j\neq i} y_{ji} p_j$
is the steady state probability of a single random walker
moving from site $i$ to $j$  with probability $y_{ij}$  \cite{MRE00,EH05}.

To make the connection between the forms
(\ref{zrpag}) and  (\ref{condf1}) we identify
\begin{equation}
x_{ij}(1) = y_{ij}p_i\quad\mbox{and }
\quad w_i(m_i) = 
\frac{p_i v_{ii}(m_i-1)}{v_{ii}(m_i)}\;.
\end{equation}
Inverting the latter equality to express $v_{ii}(m_i)$
in terms of $w_i$ yields, by virtue of  (\ref{wrs2}),
the single-site weight (\ref{zrpagf}).
Finally the consistency condition
(\ref{consrs1}) becomes
\begin{equation}
\sum_{j\ne i} x_{ij}(1)v_{ii}(m-1) =  \sum_{j\ne i} x_{ji}(1)v_{ii}(m-1),
\label{consrs2}
\end{equation}
and is satisfied  with $x_{ij}(1) = y_{ij}p_i$ since we have
\begin{equation}
\sum_{j\neq i} x_{ij}(1)=
\sum_{j\neq i} x_{ji}(1)\;.
\end{equation}

\section{Conclusion}
\label{sec:conc}
In this work we have derived the sufficent condition (\ref{phi}) along
with a consistency condition (\ref{cond}) for factorisation of the
general (discrete time, continuous mass) mass transport model on an
arbitrary graph.  In this case the single-site weight is given by
(\ref{f}). We gave in section \ref{sec:spec} specific examples of
geometries where the additional consistency condition associated with
the sufficient condition is automatically fulfilled. Moreover on a
complete graph with permutationally invariant chipping functions we
showed that the sufficent condition is, in fact, also necessary.
 
Of course a significant improvement would be to generalize condition 
(\ref{phi},\ref{cond}) to a necessary and sufficient condition. To illustrate
the challenges involved in accomplishing this task, we offer another simple
example where we can derive a condition that is both necessary and sufficient.

This example, a seemingly trivial generalization of the one in \cite
{EMZ04,ZEM04}, involves a one-dimensional lattice with chipping to both
nearest neighbour sites 
and we  also assume that the chipping kernel is translationally invariant
($l=1$ in Example 1 of section~\ref{sec:spec}).
Then, due to  the translational invariance,  (\ref{LT}) becomes 
\begin{equation}
\prod_i g(s_i)=\prod_i X(s_{i-1},s_i,s_{i+1})\;.  \label{nesex}
\end{equation}
Using a similar procedure to that employed in Example 4 of section~\ref
{sec:spec} --- taking the logarithm of (\ref{nesex}) then successive
derivatives with respect to $s_i$ and $s_j$ --- we find that the general
solution to (\ref{nesex}) is 
\begin{equation}
X(s_{i-1},s_i,s_{i+1})=K_{ii-1}(s_{i-1})K_{ii}(s_i)K_{ii+1}(s_{i+1})\frac{%
H(s_{i-1},s_i)}{H(s_i,s_{i+1})}\;.  \label{exgen}
\end{equation}
where $K_{ii-1}$, $K_{ii}$, $K_{ii+1}$
and $H$ are arbitrary functions, independent of $i$. Inverting the
Laplace transforms to give $\mathcal{P}(\mu _{ii-1},\mu _{ii},\mu _{ii+1})$
would yield a considerably more complicated form for the chipping kernel $%
\phi (\mu _{ii-1},\mu _{ii},\mu _{ii+1})$ than (\ref{phi}). In principle,
therefore, there could be a whole family of chipping kernels, generated by
the choice of $H$, which give rise to the same steady state (i.e. the same single-site
weight  $f(m)$) as that for 
$H=1$. Thus, $H$ could be thought of as  a ``gauge function''. 

In addition, for each choice of $H$, one has to ensure the consistency
condition, namely that
\begin{equation}
\phi(\mu_{i i-1}, \mu_{ii}, \mu_{i i+1})= 
\frac{{\mathcal P}(\mu_{i i-1}, \mu_{ii}, \mu_{i i+1})}{f(m_i)}
\delta(m_i- (\mu_{ii-1}+\mu_{ii}+\mu_{i i+1}))
\label{phiex}
\end{equation}
is correctly normalized to unity.
On the face of it, it may appear that this imposes a formidable constraint
on the choice of $H$. However, we now show that this consistency condition
does not impose any additional constraint on $H$. In other words, if the consistency 
condition is ensured for $H=1$, then it is automatically satisfied for 
all other choices of $H$. To see this, consider the expression of the
chipping kernel  (\ref{phiex}).
Now, the denominator $f(m_i)$ is, of course, independent of the
choice of $H$. So, to prove that the consistency condition of normalization
of $\phi$ does not impose any additional constraint on $H$, one has to 
prove that the integral
\begin{equation}
I\left( m\right) \equiv \int\D \mu_{ii-1}\D\mu_{ii}\D\mu_{ii+1} {\cal P}(\mu _{ii-1},\mu
_{ii},\mu _{ii+1})\delta \left( m-\mu _{ii-1}-\mu _{ii}-\mu _{ii+1}\right)
\label{H1}
\end{equation}
is independent of the choice of $H$. Now, taking the Laplace transform
of Eq. (\ref{H1}) one  obtains
\begin{eqnarray}
\int \D m e^{-sm} I\left( m\right)  &=&\int \D\mu_{ii-1}\D\mu_{ii}\D\mu_{ii+1} e^{-s\left( \mu
_{ii-1}+\mu _{ii}+\mu _{ii+1}\right) }{\cal P} (\mu _{ii-1},\mu _{ii},\mu
_{ii+1}) \nonumber \\
&=&X(s,s,s) \nonumber \\
&=&K_{ii-1}(s)K_{ii}(s)K_{ii+1}(s)
\end{eqnarray}
where we have used definition (\ref{LT}) in going from  the first line to the second
and Eq. (\ref{exgen}) in going from the second line to the 
third. But this expression on the rhs does not contain the ``gauge function'' $H$ 
thereby proving that the integral of ${\cal P}$ is independent of the choice of $H$.
Thus the consistency condition is automatically fulfilled for arbitrary choices of $H$
as long as it is ensured for $H=1$, which is indeed the case as shown in 
Sec.~\ref{sec:spec} Example 1.

A more serious constraint on $H$ is that the inverse Laplace
transform of (\ref{exgen}) must be \emph{non-negative}. The case 
$H=1$ in (\ref{exgen}) obviously imposes the trivial constraint that the
inverse Laplace transform of the $K$'s  be non-negative.
It remains to be shown
whether there is also   a 
class of $H \neq 1$  which lead to valid chipping kernels. 
If it could be shown that this class
is non-empty, the one-dimensional $l=1$ example we have discussed
would show  explicitly that it is 
\emph{not} necessary for $\phi $ to be of the form (\ref{phi}).  
Mapping the extent of
this class of $H$ (if indeed it is non-empty) remains a daunting task. 

For an arbitrary graph, we have made some in-roads with similar
considerations. Unfortunately, the most general solution to (\ref{LT})
remains difficult to formulate. 

Finally we note that having a factorised steady state opens the door for the
study of condensation as in \cite{MEZ05,longp}. Thus one should be able to
analyse condensation in various geometries or even on scale-free networks 
\cite{NSL}.

\ack
 RKPZ acknowledges the support
of the US National Science Foundation through DMR-0414122.

\end{document}